\documentclass[jmse,article,submit,pdftex,moreauthors]{Definitions/mdpi}
\renewcommand{\linenumbers}{}
\firstpage{1}
\pubvolume{1}
\issuenum{1}
\articlenumber{0}
\pubyear{2023}
\copyrightyear{2023}
\datereceived{ }
\daterevised{ } 
\dateaccepted{ }
\datepublished{ }
\hreflink{https://doi.org/} 

\Title{Variational Principles for Nonbarotropic Fluid Dynamics}

\TitleCitation{Variational Principles for Nonbarotropic Fluid Dynamics}

\Author{Asher Yahalom$^{1,2}$\orcidA{}}

\AuthorNames{Asher Yahalom}

\AuthorCitation{Yahalom, A.}

\address{%
$^{1}$ \quad Ariel University, Faulty of Engineering, Department of Electrical \& Electronic Engineering, Ariel 40700, Israel; asya@ariel.ac.il\\
$^{2}$ \quad Ariel University, Center for Astrophysics, Geophysics, and Space Sciences (AGASS),
Ariel 40700, Israel;}

\corres{Correspondence: asya@ariel.ac.il; Tel. +972-54-7740294: }

\abstract{Barotropic fluid flows with the same circulation structure as steady flows generically have comoving physical surfaces on which the vortex lines lie. These become Bernoullian surfaces when the flow is steady. When these surfaces are nested (vortex line foliation) with the topology of cylinders, toroids or a combination of both, a Clebsch representation of the flow velocity can be introduced. This is then used to reduce the number of functions to be varied in the variational principles for such flows. Here we extend the work to non barotropic flows and study the implication for variational analysis and conserved quantities of topological significance such as circulation and helicity.}


\begin{document}

\newcommand{\beq} {\begin{equation}}
\newcommand{\enq} {\end{equation}}
\newcommand{\ber} {\begin {eqnarray}}
\newcommand{\enr} {\end {eqnarray}}
\newcommand{\eq} {equation}
\newcommand{\eqs} {equations }
\newcommand{\mn}  {{\mu \nu}}
\newcommand{\abp}  {{\alpha \beta}}
\newcommand{\ab}  {{\alpha \beta}}
\newcommand{\sn}  {{\sigma \nu}}
\newcommand{\rhm}  {{\rho \mu}}
\newcommand{\sr}  {{\sigma \rho}}
\newcommand{\bh}  {{\bar h}}
\newcommand{\br}  {{\bar r}}
\newcommand {\er}[1] {equation (\ref{#1}) }
\newcommand {\ers}[1] {equations (\ref{#1}) }
\newcommand {\ern}[1] {equation (\ref{#1})}
\newcommand {\Ern}[1] {Equation (\ref{#1})}
\newcommand{\hdz}  {\frac{1}{2} \Delta z}
\newcommand{\curl}[1]{\vec{\nabla} \times #1} 
\newcommand {\Sc} {Schr\"{o}dinger}
\newcommand {\SE} {Schr\"{o}dinger equation }
\newcommand{\ce}  {continuity equation }
\newcommand{\va}  {\vec \alpha}

\section{Introduction}

A comprehensive introduction to the subject of variational principles of fluid dynamics
is given in \cite{AYLB} and will not be repeated here, the interested reader is referred to
the original paper.

In this paper we extend what it known on variational barotropic fluid dynamics to non-barotropic fluid dynamics, this has also impact on the form and validity of conservation laws
such as the conservation of circulation and the conservation of cross helicity which has a well known topological interpretation in terms of the knottiness of vortex lines \cite{Moff2}.

Non-barotropic flows are distinguished from Barotropic flows by their more realistic equation of
state. The internal energy of those flows and therefore the pressure, depend on both the density
and specific entropy in contrast to the (over)simplified equation of state of a Barotropic flow which is a function of density alone. This allows us to discuss the effect of entropy and temperature on the dynamics of the flow and points the way to further developments towards the variational analysis of non ideal flows in which heat losses play an important rule.

We start by introducing the basic equations, this will be followed by a discussion of
the Lagrangian variational principle. Then we discuss the Eulerian variational principle
and its simplification, including its stationary form. This is followed with the analysis of conservation laws within the Non-Barotropic framework. Finally we demonstrate that
the non-barotropic variational problem can be formulated in terms of only four functions.

\section{Basic Equations of non-Stationary Non-Barotropic Fluid Dynamics}

Non-Barotropic Eulerian fluids can be described in terms of five functions the
velocity $\vec v$, density $\rho$, and specific entropy $s$. Those functions need to satisfy the
 continuity, Euler equations and the condition of ideal flows which is the lack of heat
 losses:
\beq
\frac{\partial{\rho}}{\partial t} + \vec \nabla \cdot (\rho \vec v ) = 0
\label{massconb}
\enq
\beq
\frac{d \vec v}{d t} \equiv
\frac{\partial \vec v}{\partial t}+(\vec v \cdot \vec \nabla)\vec v  = -\frac{\vec \nabla p (\rho,s)}{\rho} - \vec \nabla \Psi
\label{Eulerb}
\enq
\beq
\frac{d s}{d t} \equiv \frac{\partial s}{\partial t}+(\vec v \cdot \vec \nabla) s  = 0
\label{seq}
\enq
In which the pressure $p (\rho,s)$ is assumed to be a given function of the density and specific entropy $s$, $\Psi$ is some specific force potential (which can be gravitational or electromagnetic), $\frac{\partial }{\partial t}$ is a partial temporal derivative, $\vec \nabla$ has its usual meaning in vector analysis and $\frac{d }{d t}$
is the material temporal derivative. \Ern{massconb} tells us that the mass of each fluid element is conserved, \ern{Eulerb} is just Newton's second law for continuous matter, while \ern{seq} is a
mathematical expression for the fact that in an ideal flow, heat is not conducted, nor created but
only convected.

\section{Thermodynamics \& Vortex dynamics}

Taking the curl of \ern{Eulerb} will lead to:
\beq
\frac{\partial{\vec \omega}}{\partial t} = \vec \nabla \times (\vec v \times \vec \omega)
- \vec \nabla \times \left(\frac{\vec \nabla p}{\rho}\right)
\label{omegeq0}
\enq
in which:
\beq
\vec \omega = \vec \nabla \times \vec v
\label{vortic0}
\enq
is the vorticity.
Let us now look at the thermodynamics of the fluid.  The fluid is made of "fluid elements" \cite{Eckart,Bertherton}, practically a "fluid element" is a point particle which has an infinitesimal mass $d M_{\vec \alpha}$, position vector $\vec x (\vec \alpha,t)$ and velocity $\vec v (\vec \alpha,t) \equiv \frac{d \vec x (\vec \alpha,t)}{dt}$.
As the "fluid element" is not truly a point particle it has also an infinitesimal volume $d V_{\vec \alpha}$, infinitesimal entropy $d S_{\vec \alpha}$, and an infinitesimal internal energy $d E_{in~\vec \alpha}$.
 It is customary to define densities for the Lagrangian, mass and charge of every fluid element as
 follows:
 \beq
  \rho_{\va} \equiv \frac{d M_{\va}}{d V_{\va}}.
\label{dens}
\enq
The density can be thought of as a function of the location $\vec x$, where the "fluid element" labelled $\va$ happens to be in time $t$, for example:
\beq
 \rho (\vec x, t) \equiv \rho (\vec x (\va,t), t) \equiv \rho_{\va} (t)
\label{dens2}
\enq
It is also customary to define the specific internal energy $\varepsilon_{\va}$ as follows:
\beq
 \varepsilon_{\va} \equiv \frac{d E_{in~\va}}{d M_{\va}} \quad \Rightarrow  \quad
 \rho_{\va} \varepsilon_{\va} = \frac{d M_{\va}}{d V_{\va}} \frac{d E_{in~\va}}{d M_{\va}} =
 \frac{d E_{in~\va}}{d V_{\va}}
\label{specifint}
\enq
which will be important for later sections of the current paper. In an ideal fluid the "fluid element" does not exchange mass nor heat with other fluid elements, so it follows that:
\beq
\Delta d M_{\vec \alpha} = \Delta d S_{\vec \alpha}  =0.
\label{consvl}
\enq
in which we use the $\Delta$ symbol to denote change.
Moreover, according to thermodynamics a change in the internal energy of a "fluid element" satisfies
the equation:
\beq
\Delta d E_{in~\vec \alpha} = T_{\va} \Delta dS_{\vec \alpha} - p_{\va} \Delta dV_{\vec \alpha},
\label{thermo}
\enq
the first term describes the heating energy gained by the "fluid element" while the second terms describes the work done by the "fluid element" on neighbouring elements. $T_{\va}$ is the temperature of the "fluid element" and $p_{\va}$ is the pressure of the same. As the mass of the fluid element does not change we may divide the above expression by $d M_{\vec \alpha}$ to obtain
the variation of the specific energy as follows:
\ber
\Delta \varepsilon_{\va}  &=& \Delta\frac{d E_{in~\vec \alpha}}{d M_{\va}}
 =  T_{\va} \Delta\frac{ dS_{\vec \alpha}}{d M_{\va}} - p_{\va} \Delta \frac{dV_{\vec \alpha}}{d M_{\va}}
 \nonumber \\
 &=& T_{\va} \Delta s_{\va} - p_{\va} \Delta \frac{1}{\rho_{\va}}
 = T_{\va} \Delta s_{\va} + \frac{p_{\va}}{\rho_{\va}^2} \Delta \rho_{\va}. \qquad
 s_{\va} \equiv \frac{ dS_{\vec \alpha}}{d M_{\va}}
\label{thermo2b}
\enr
in which $s_{\va}$ is the specific entropy of the fluid element. It follows that:
\beq
\frac{\partial \varepsilon }{\partial s} = T, \qquad
\frac{\partial \varepsilon }{\partial \rho} = \frac{p}{\rho^2}.
\label{thermo2c}
\enq
Another important thermodynamic quantity that we will use later is the Enthalpy defined for
a fluid element as:
\beq
 dW_{\va} = d E_{in~\vec \alpha} + p_{\va} d V_{\va}.
\label{thermo2d}
\enq
and the specific enthalpy:
\beq
 w_{\va} =\frac{dW_{\va}}{d M_{\va}}= \frac{dE_{in~\vec \alpha}}{d M_{\va}}
 +p_{\va} \frac{dV_{\va}}{d M_{\va}} = \varepsilon_{\va} + \frac{p_{\va}}{\rho_{\va}}.
\label{thermo2e}
\enq
Combining the above result with \ern{thermo2c} it follows that:
\beq
 w = \varepsilon + \frac{p}{\rho} = \varepsilon + \rho \frac{\partial \varepsilon }{\partial \rho}
 = \frac{\partial (\rho \varepsilon) }{\partial \rho}.
\label{thermo2f}
\enq
Moreover:
\beq
 \frac{\partial w }{\partial \rho} = \frac{\partial (\varepsilon + \frac{p}{\rho})}{\partial \rho}
  = - \frac{p}{\rho^2} + \frac{1}{\rho} \frac{\partial p }{\partial \rho}+ \frac{\partial \varepsilon }{\partial \rho} =
  - \frac{p}{\rho^2} + \frac{1}{\rho} \frac{\partial p }{\partial \rho}+ \frac{P}{\rho^2} =
   \frac{1}{\rho} \frac{\partial p }{\partial \rho}.
\label{thermo2g}
\enq
Now we can look again at the gradient of the pressure appearing in \ern{omegeq0}:
\beq
\frac{\vec \nabla p}{\rho} = \vec \nabla \left( \frac{p}{\rho} \right) + \frac{p}{\rho^2}
\vec \nabla \rho = \vec \nabla (w - \varepsilon) + \frac{p}{\rho^2}
\vec \nabla \rho
\label{gradp}
\enq
in which we have used \ern{thermo2f} in the above derivation. Looking again at \ern{thermo2b}
and considering a change of the fluid element location, leads to the following identity for
the specific internal energy gradient:
\beq
 \vec \nabla \varepsilon = T \vec \nabla s + \frac{p}{\rho^2} \vec \nabla \rho.
\label{gradvareps}
\enq
Combining \ern{gradp} with \ern{gradvareps} leads to the equation:
\beq
\frac{\vec \nabla p}{\rho}  = \vec \nabla w - T \vec \nabla s
\label{gradp2}
\enq
We can now plug the above equation into \ern{omegeq0} to obtain:
\beq
\frac{\partial{\vec \omega}}{\partial t} = \vec \nabla \times (\vec v \times \vec \omega)
+ \vec \nabla T \times \vec \nabla s.
\label{omegeq1}
\enq
Equation (\ref{omegeq0}) describes the fact that the vorticity lines are "frozen"
within the flow if the temperature or specific entropy are uniform or their gradients are parallel.
The fact that vorticity lines are "frozen" (or not) is obviously connected to the conservation of
circulation and helicity and thus signifies the profound connection in flows between thermodynamics
and vortex dynamics.

\section{The Lagrangian Variational Approach}

The action and Lagrangian for each "fluid element" are similar to that of a point particle,
with one difference: its internal energy. The variational formalism of point particles is
described in appendix \ref{pointpart}. Thus according to \ern{classparticlej} we obtain
the expression:
\ber
 d {\cal A}_{\va} &=& \int_{t1}^{t2} d L_{\va} dt,
 \nonumber \\
 d L_{\va} &\equiv&
 \frac{1}{2} d M_{\va}~ v(\va,t)^2 - d M_{\va} \Psi_{\va}  - d E_{in~\vec \alpha}.
 \label{classparticleactcf}
 \enr
in which we replace the discrete index $j$ with the continuous index $\vec \alpha$.
Of course, all the above quantities are calculated for a specific value of the label $\vec \alpha$, while the action and Lagrangian of the entire fluid, should be summed (or integrated) over all possible $\vec \alpha$'s. That is:
\ber
L &=& \int_{\va}  d L_{\vec \alpha}
\nonumber \\
{\cal A} &=& \int_{\va} d {\cal A}_{\vec \alpha} = \int_{t1}^{t2} \int_{\va}  d L_{\vec \alpha} dt
 =  \int_{t1}^{t2} L dt.
 \label{fluac1}
 \enr
Thus we can write the following equations for the Lagrangian density:
\ber
 {\cal L}_{\va} &=& \frac{d L_{\va}}{d V_{\va}}
 \nonumber \\
 {\cal L}_{\va} &\equiv& \frac{1}{2} \rho_{\va} v (\va,t)^2 - \rho_{\va} \varepsilon_{\va}
- \rho_{\va} \Psi_{\va}.
 \label{lagdensity}
 \enr
 The above expression allows us to write the Lagrangian as a spatial integral:
 \beq
L = \int_{\va}  d L_{\vec \alpha} = \int_{\va} {\cal L}_{\va} d V_{\va}
= \int {\cal L} (\vec x,t) d^3 x
\label{Lspatint}
\enq
We shall now calculate the variation of the fluid action. The only term which is somewhat different
from the classical case is the internal energy term whose variation is given in \ern{thermo}.
As we assume an ideal fluid, there is no heat generation not heat conduction or heat radiation, and thus heat can only be moved around along the trajectory of the "fluid elements", that is only convection is taken into account. Thus $\Delta d S_{\vec \alpha}  =0$
and we have:
\beq
\Delta d E_{in~\vec \alpha} = - p_{\va} \Delta dV_{\vec \alpha}.
\label{thermo2}
\enq
Our next step would to be to evaluate the variation of the volume element. Suppose a time $t$ the
volume of the fluid element labelled by $\va$ is described as:
\beq
 dV_{\va,t} = d^3 x(\va, t)
\label{volelem}
\enq
Using the Jacobian determinant we may relate this to the same element at $t=0$:
\beq
  d^3 x(\va, t)  = J  d^3 x(\va, 0), \qquad
  J \equiv \vec \nabla_0 x_1 \cdot (\vec \nabla_0 x_2 \times \vec \nabla_0 x_3)
\label{volelem2}
\enq
In which $\vec \nabla_0$ is taken with respect to the coordinates of the fluid elements at $t=0$: $\vec \nabla_0 \equiv (\frac{\partial }{\partial x(\va,0)_1},\frac{\partial }{\partial x(\va,0)_2},\frac{\partial }{\partial x(\va,0)_3})$. As both the actual and varied "fluid element" trajectories start at the same point it follows that:
\ber
\Delta dV_{\va,t} &=& \Delta d^3 x(\va, t)  = \Delta J ~ d^3 x(\va, 0)  = \frac{\Delta J}{J} d^3 x(\va, t) = \frac{\Delta J}{J}  dV_{\va,t},
\nonumber \\
(\Delta d^3 x(\va, 0) &=& 0).
\label{volelem3}
\enr
The variation of $J$ can be easily calculated as:
\beq
    \Delta J = \vec \nabla_0 \Delta x_1 \cdot (\vec \nabla_0 x_2 \times \vec \nabla_0 x_3)
    + \vec \nabla_0  x_1 \cdot (\vec \nabla_0 \Delta x_2 \times \vec \nabla_0 x_3)
    + \vec \nabla_0  x_1 \cdot (\vec \nabla_0  x_2 \times \vec \nabla_0 \Delta x_3),
\label{volelem4}
\enq
Introducing the notation $\vec \xi \equiv \Delta \vec x$ it follows that:
\ber
& & \vec \nabla_0 \Delta x_1 \cdot (\vec \nabla_0 x_2 \times \vec \nabla_0 x_3)
= \vec \nabla_0 \xi_1 \cdot (\vec \nabla_0 x_2 \times \vec \nabla_0 x_3)
\nonumber \\
&=& \partial_k \xi_1 \vec \nabla_0 x_k \cdot (\vec \nabla_0 x_2 \times \vec \nabla_0 x_3)
= \partial_1 \xi_1 \vec \nabla_0 x_1 \cdot (\vec \nabla_0 x_2 \times \vec \nabla_0 x_3)=
\partial_1 \xi_1 J.
\nonumber \\
& & \vec \nabla_0  x_1 \cdot (\vec \nabla_0 \Delta x_2 \times \vec \nabla_0 x_3)
= \vec \nabla_0 x_1 \cdot (\vec \nabla_0 \xi_2 \times \vec \nabla_0 x_3)
\nonumber \\
&=& \partial_k \xi_2 \vec \nabla_0 x_1 \cdot (\vec \nabla_0 x_k \times \vec \nabla_0 x_3)
= \partial_2 \xi_2 \vec \nabla_0 x_1 \cdot (\vec \nabla_0 x_2 \times \vec \nabla_0 x_3)=
\partial_2 \xi_2 J.
\nonumber \\
& & \vec \nabla_0  x_1 \cdot (\vec \nabla_0  x_2 \times \vec \nabla_0 \Delta x_3)
= \vec \nabla_0 x_1 \cdot (\vec \nabla_0 x_2 \times \vec \nabla_0 \xi_3)
\nonumber \\
&=& \partial_k \xi_3 \vec \nabla_0 x_1 \cdot (\vec \nabla_0 x_2 \times \vec \nabla_0 x_k)
= \partial_3 \xi_3 \vec \nabla_0 x_1 \cdot (\vec \nabla_0 x_2 \times \vec \nabla_0 x_3)=
\partial_3 \xi_3 J.
\label{volelem5}
\enr
Combining the above results, it follows that:
\beq
    \Delta J = \partial_1 \xi_1 J + \partial_2 \xi_2 J + \partial_3 \xi_3 J =
     \vec \nabla \cdot \vec \xi~ J.
\label{volelem6}
\enq
Which allows us to calculate the variation of the volume of the "fluid element":
\beq
\Delta dV_{\va,t}  = \vec \nabla \cdot \vec \xi ~ dV_{\va,t}.
\label{volelem7}
\enq
And thus the variation of the internal energy is:
\beq
\Delta d E_{in~\vec \alpha} = - p \vec \nabla \cdot \vec \xi ~ dV_{\va,t}.
\label{thermo3}
\enq
The internal energy is the only novel element with respect to the single particle scenario and system of particles scenario described in the appendix, thus the rest of the variation analysis is straight forward. Varying \ern{classparticleactcf} we obtain:
\ber
 \Delta d {\cal A}_{\va} &=& \int_{t1}^{t2} \Delta d L_{\va} dt,
 \nonumber \\
\Delta d L_{\va} &=&  d M_{\va}~\left(\vec v(\va,t) \cdot \Delta \vec v(\va,t)-\Psi_{\va}\right)   - \Delta d E_{in~\vec \alpha} .
 \label{varclassparticleactcf}
 \enr
Notice that:
\beq
\Delta \vec v(\va,t) =  \Delta \frac{d \vec x (\vec \alpha,t)}{dt} = \frac{d \Delta \vec x (\vec \alpha,t)}{dt} = \frac{d \vec \xi (\vec \alpha,t)}{dt}.
\label{vvar}
\enq
After some steps which are described prior to \ern{delLf} we obtain the variation of $d L_{\va}$:
\beq
\Delta d L_{\va} =
 \frac{d (d M_{\va} \vec v_{\va} \cdot \vec \xi_{\va})}{dt} - d M_{\va}
 \left(\frac{d {\vec v}_{\va} }{dt}+ \vec \nabla \Psi_{\va}\right) \cdot \vec \xi_{\va} + p \vec \nabla \cdot \vec \xi_{\va} ~ dV_{\va,t}.
 \label{delLffl}
 \enq
 The variation of the action of a single fluid element is thus:
 \ber
  \Delta d {\cal A}_{\va} &=& \int_{t1}^{t2} \Delta dL_{\va} dt =
  \left. d M_{\va} \vec v (\va,t) \cdot \vec \xi_{\va} \right|_{t1}^{t2}
  \nonumber \\
   &-& \int_{t1}^{t2}(d M_{\va} \left(\frac{d {\vec v}_{\va} }{dt}+ \vec \nabla \Psi_{\va}\right) \cdot \vec \xi_{\va} - p \vec \nabla \cdot \vec \xi_{\va} ~ dV_{\va,t} )  dt.
 \label{delA1fl}
 \enr
The variation of the total action of the fluid is thus:
\ber
\Delta {\cal A} &=& \int_{\va} d {\cal A}_{\vec \alpha} =
\left. \int_{\va} d M_{\va} \vec v (\va,t)  \cdot \vec \xi_{\va} \right|_{t1}^{t2}
  \nonumber \\
   &-& \int_{t1}^{t2}\int_{\va} (d M_{\va} \left(\frac{d {\vec v}_{\va} }{dt}+ \vec \nabla \Psi_{\va}\right) \cdot \vec \xi_{\va} - p \vec \nabla \cdot \vec \xi_{\va} ~ dV_{\va} )  dt.
 \label{varAfl2}
 \enr
 Now according to \ern{dens} we may write:
 \beq
 d M_{\va} = \rho_{\va}~dV_{\va},
 \label{dvq}
 \enq
using the above relations we may convert the $\va$ integral into a volume integral and thus write
the variation of the fluid action in which we suppress the $\va$ labels:
\beq
\Delta {\cal A} =
\left. \int \rho \vec v   \cdot \vec \xi dV \right|_{t1}^{t2}
  - \int_{t1}^{t2}\int (\rho (\frac{d \vec v}{dt} + \vec \nabla \Psi) \cdot \vec \xi - p \vec \nabla \cdot \vec \xi)  dV  dt.
 \label{varAfl3}
 \enq
Now, since:
\beq
p \vec \nabla \cdot \vec \xi = \vec \nabla \cdot (p \vec \xi) - \vec \xi \cdot \vec \nabla p,
\label{Pxi}
 \enq
and using Gauss theorem the variation of the action can be written as:
\ber
\Delta {\cal A} &=&
\left. \int \rho \vec v  \cdot \vec \xi dV \right|_{t1}^{t2}
 \nonumber \\
  &-& \int_{t1}^{t2}\left[\int (\rho (\frac{d \vec v}{dt}+ \vec \nabla \Psi) + \vec \nabla p)\cdot \vec \xi   dV  - \oint p \vec \xi \cdot d \vec \Sigma \right] dt.
 \label{varAfl4}
 \enr
 It follows that the variation of the action will vanish for a $\vec \xi$ such that $\vec \xi (t1) = \vec \xi (t2)=0$ and vanishing on a surface encapsulating the fluid, but other than that arbitrary only if the Euler equation for a charged fluid is satisfied, that is:
 \beq
 \frac{d \vec v}{dt}= -\frac{\vec \nabla p}{\rho} -\vec \nabla \Psi, \qquad
 \frac{d }{dt} = \frac{\partial }{\partial t} + \vec v \cdot \vec \nabla
 \label{Eul1}
 \enq
  thus except from the pressure terms the equation is similar to that of a point particle. In experimental fluid dynamics it is more convenient to describe a fluid in terms of quantities at
 a specific location, rather than quantities associated with unseen infinitesimal "fluid elements". This road leads to the Eulerian description of fluid dynamics and thinking in terms of flow fields rather than in terms of a velocity of "fluid elements" as will be discussed in the section \ref{Eulflu}.

\section{The Eulerian Variational Principle}
\label{Eulflu}

The Eulerian approach is radically differen way to study flows. Instead of fluid elements
we look at fluid fields which may be scalar, such as the mass density $\rho(\vec x,t)$ and
the specific entropy $s(\vec x,t)$ or vector such as the velocity field $\vec v(\vec x,t)$.
To this we will need to add auxiliary functions for maintaining information that is lost in
the transformation from the Lagrangian to the Eulerian picture, such as the particles mass and
identity. Here we follow in the footsteps of Clebsch \cite{Clebsch1,Clebsch2},
Davydov \cite{Davidov} and  Seliger \& Whitham \cite{Seliger}.
This will serve as a starting point for the next section in which we will show how the variational principle can be simplified further. Consider the action:
\ber
A & \equiv & \int {\cal L} d^3 x dt
\nonumber \\
{\cal L} & \equiv & {\cal L}_1 + {\cal L}_2
\nonumber \\
{\cal L}_1 & \equiv & \rho (\frac{1}{2} \vec v^2 - \varepsilon (\rho,s)-\Psi), \qquad
{\cal L}_2 \equiv  \nu [\frac{\partial{\rho}}{\partial t} + \vec \nabla \cdot (\rho \vec v )]
- \rho \alpha \frac{d \beta}{dt} - \rho \sigma \frac{d s}{dt}
\label{Lagactionsimpb}
\enr
in the above ${\cal L}_1$ is simply the Lagrangian density given in \ern{lagdensity}.
${\cal L}_2$ is composed from a set of constraints which are enforced by the Lagrange multipliers
$\nu,\alpha,\sigma$. The $\nu,\alpha$ functions appear also in the barotropic variational
formalism \cite{AYLB}, however, the $\sigma$ Lagrange multiplier is of course unique to non-barotropic flows and appears also in non-barotropic magnetohydrodynamics \cite{nonBarotropic,nonBarotropic2}.
Variation with respect to $\nu,\alpha,\sigma$ will obviously yield the following \eqs:
\ber
& & \frac{\partial{\rho}}{\partial t} + \vec \nabla \cdot (\rho \vec v ) = 0
\nonumber \\
& & \rho \frac{d \beta}{dt} = 0
\nonumber \\
& & \rho \frac{d s}{dt} = 0
\label{lagmulb}
\enr
Provided $\rho$ is not null those are just the continuity \ern{massconb} and
the conditions that $\beta$ and $s$ are comoving, that is they do not change for any flow element.
Let us take an arbitrary variational derivative of the above action with
respect to $\vec v$, this will result in:
\ber
\delta_{\vec v} A & = & \int d^3 x dt \rho \delta \vec v \cdot
[\vec v - \vec \nabla \nu - \alpha \vec \nabla \beta - \sigma \vec \nabla s]
\nonumber \\
 & + & \oint d \vec A \cdot \delta \vec v \rho \nu + \int d \vec \Sigma \cdot \delta \vec v \rho [\nu],
\label{delActionvb}
\enr
the above boundary terms contain integration over the external boundary $\oint d \vec A$ and an
integral over the cut $\int d \vec \Sigma$ that must be introduced in case that $\nu$ is not single
valued, more on this case in later sections.
The external boundary term vanishes; in the case of astrophysical flows for which $\rho=0$ on the free flow boundary, or the case in which the fluid is contained in a vessel which induces a no flux boundary condition $\delta \vec v \cdot \hat n =0$
($\hat n$ is a unit vector normal to the boundary). The cut "boundary" term vanish when the velocity field varies only parallel to the cut that is it satisfies a Kutta type condition. If the boundary terms vanish  $\vec v$ must have the following form:
\beq
\vec v = \hat {\vec v} \equiv \alpha \vec \nabla \beta + \sigma \vec \nabla s + \vec \nabla \nu
\label{vformb}
\enq
this is a generalization of the Clebsch representation of the flow field (see for example \cite{Eckart,Clebsch1,Clebsch2}, \cite[page 248]{Lamb H.}).
Let us now take the variational derivative with respect to the density $\rho$, we obtain:
\ber
\delta_{\rho} A & = & \int d^3 x dt \delta \rho
[\frac{1}{2} \vec v^2 - w -\Psi  - \frac{\partial{\nu}}{\partial t} -  \vec v \cdot \vec \nabla \nu]
\nonumber \\
 & + & \oint d \vec A \cdot \vec v \delta \rho  \nu +\int d \vec \Sigma \cdot \vec v \delta \rho  [\nu] +
  \int d^3 x \nu \delta \rho |^{t_1}_{t_0}
\label{delActionrhob}
\enr
Hence provided that $\delta \rho$ vanishes on the boundary of the domain, on the cut and in initial
and final times the following \eq must be satisfied:
\beq
\frac{d \nu}{d t} = \frac{1}{2} \vec v^2 - w - \Psi
\label{nueqb}
\enq
Let us now calculate the variation with respect to the specific entropy $s$:
\ber
\delta_{s} A  & = &  \int d^3 x dt \delta s
[\frac{\partial{(\rho \sigma)}}{\partial t} +  \vec{\nabla} {\cdot} (\rho \sigma \vec{v})- \rho T]
+ \int dt \oint d \vec{A} {\cdot} \rho \sigma \vec{v}  \delta s
 \nonumber \\
 & - &  \int d^3 x \rho \sigma \delta s |^{t_1}_{t_0},
\label{delActions}
\enr
in which the temperature is $T=\frac{\partial \varepsilon}{\partial s}$ is defined in \ern{thermo2c}. We notice that according
to \ern{vformb} $\sigma$ is single valued and hence no cuts are needed. Taking into account the continuity \ern{lagmulb} we obtain for locations in which the density $\rho$ is not null the result:
\beq
\frac{d \sigma}{dt} =T,
\label{sigmaeq}
\enq
provided that $\delta_{s} A$ vanished for an arbitrary $\delta s$.

Finally we have to calculate the variation with respect to $\beta$
this will lead us to the following results:
\ber
\delta_{\beta} A & = & \int d^3 x dt \delta \beta
[\frac{\partial{(\rho \alpha)}}{\partial t} +  \vec \nabla \cdot (\rho \alpha \vec v)]
\nonumber \\
 & - & \oint d \vec A \cdot \vec v \rho \alpha \delta \beta -\int d \vec \Sigma \cdot \vec v \rho \alpha [\delta \beta]
 - \int d^3 x \rho \alpha \delta \beta |^{t_1}_{t_0}
\label{delActionchib}
\enr
Hence choosing $\delta \beta$ in such a way that the temporal and
spatial boundary terms vanish (this includes choosing $\delta \beta$ to be continuous on the cut if one needs to introduce such a cut) in the above integral will lead to
the equation:
\beq
\frac{\partial{(\rho \alpha)}}{\partial t} +  \vec \nabla \cdot (\rho \alpha \vec v) =0
\enq
Using the continuity \ern{massconb} this will lead to the equation:
\beq
\rho \frac{d \alpha}{dt} = 0
\label{alphacon}
\enq
Hence for $\rho \neq 0$ both $\alpha$ and $\beta$ are comoving coordinates. Since
the vorticity can be easily calculated from \er{vformb} to be:
\beq
\vec \omega = \vec \nabla \times \vec v =  \vec \nabla \alpha \times \vec \nabla \beta
+ \vec \nabla \sigma \times \vec \nabla s,
\label{vorticb}
\enq
Calculating $\frac{\partial{\vec \omega}}{\partial t}$ in which $\omega$ is
given by \ern{vorticb} and taking into
account \ern{alphacon}, \ern{sigmaeq} and \ern{lagmulb} will yield \ern{omegeq1}.

\section{Euler's equations}
\label{Eulerequations}

Although we obtained from the variational principle the continuity  and specific entropy
conservation equations (\ref{massconb},\ref{seq}) the rest of the variational equations seem unrelated to the Euler equations (\ref{Eulerb}), however, this impression is false.

We shall now show that a velocity field given by \ern{vformb}, such that the
functions $\alpha, \beta, \nu, \sigma, s$ satisfy the corresponding equations
(\ref{lagmulb},\ref{nueqb},\ref{sigmaeq},\ref{alphacon}) must satisfy Euler's equations.
Let us calculate the material derivative of $\vec v$ using \ern{sigmaeq} and \ern{alphacon}:
\beq
\frac{d\vec v}{dt} = \frac{d\vec \nabla \nu}{dt}  + \frac{d\alpha}{dt} \vec \nabla \beta +
 \alpha \frac{d\vec \nabla \beta}{dt}+ \frac{d \sigma}{dt} \vec \nabla s +
 \sigma \frac{d\vec \nabla s}{dt}
 = \frac{d\vec \nabla \nu}{dt} +  \alpha \frac{d\vec \nabla \beta}{dt}+ T \vec \nabla s +
 \sigma \frac{d\vec \nabla s}{dt}
\label{dvform12b}
\enq
It can be easily shown using \ern{lagmulb} and \ern{nueqb} that:
\ber
\frac{d\vec \nabla \nu}{dt} & = & \vec \nabla \frac{d \nu}{dt}- \vec \nabla v_k \frac{\partial \nu}{\partial x_k}
 = \vec \nabla (\frac{1}{2} \vec v^2 - w -\Psi)- \vec \nabla v_k \frac{\partial \nu}{\partial x_k}
 \nonumber \\
 \frac{d\vec \nabla \beta}{dt} & = & \vec \nabla \frac{d \beta}{dt}- \vec \nabla v_k \frac{\partial \beta}{\partial x_k}
 = - \vec \nabla v_k \frac{\partial \beta}{\partial x_k}
 \nonumber \\
 \frac{d\vec \nabla s}{dt} & = & \vec \nabla \frac{d s}{dt}- \vec \nabla v_k \frac{\partial s}{\partial x_k}
 = - \vec \nabla v_k \frac{\partial s}{\partial x_k}
  \label{dnablab}
\enr
In which $x_k$ is a Cartesian coordinate and a summation convention is assumed. Inserting the result from equations (\ref{dnablab}) into \ern{dvform12b} yields:
\ber
\frac{d\vec v}{dt} &=& - \vec \nabla v_k (\frac{\partial \nu}{\partial x_k} +
 \alpha \frac{\partial \beta}{\partial x_k} +  \sigma \frac{\partial s}{\partial x_k}  ) + \vec \nabla (\frac{1}{2} \vec v^2 -w -\Psi) + T \vec \nabla s
 \nonumber \\
&=& - \vec \nabla v_k v_k + \vec \nabla (\frac{1}{2} \vec v^2 - w -\Psi) + T \vec \nabla s
 = - \frac{\vec \nabla p}{\rho}- \vec \nabla \Psi,
\label{dvform2bb}
\enr
in which we have used both \ern{vformb} and \ern{gradp2}.
 This proves that the Euler equations can be derived from the action given in \ern{Lagactionsimpb} and hence all the equations of  fluid dynamics can be derived from the above action
without restricting the variations in any way. Taking the curl of \ern{dvform2bb} will lead to \ern{omegeq1}.

\section{Simplified Eulerian action}
\label{simpact}

The reader of this paper might argue that the authors have introduced unnecessary complications
to the theory of fluid dynamics by adding four more functions $\alpha,\beta,\nu, \sigma$ to the standard set $\vec v,\rho,s$. In the following we will show that this is not so and the action given in \ern{Lagactionsimpb} in
a form suitable for a pedagogic presentation can be simplified with respect to the number of unknown functions. It is easy to show
that the Lagrangian density appearing in \ern{Lagactionsimpb} can be written in the form:
\ber
{\cal L} & = & -\rho [\frac{\partial{\nu}}{\partial t} + \alpha \frac{\partial{\beta}}{\partial t}
+ \sigma \frac{\partial{s}}{\partial t} + \varepsilon (\rho,s) + \Psi] +
\frac{1}{2}\rho [(\vec v-\hat{\vec v})^2-\hat{\vec v}^2]
\nonumber \\
& + &  \frac{\partial{(\nu \rho)}}{\partial t} + \vec \nabla \cdot (\nu \rho \vec v )
\label{Lagactionsimpb4}
\enr
In which $\hat{\vec v}$ is a shorthand notation for $\vec \nabla \nu + \alpha \vec \nabla \beta
+ \sigma \vec \nabla s $ (see \ern{vformb}). Thus ${\cal L}$ has three contributions:
\ber
{\cal L} & = & \hat {\cal L} + {\cal L}_{\vec v}+ {\cal L}_{boundary}
\nonumber \\
\hat {\cal L} &\equiv & -\rho [\frac{\partial{\nu}}{\partial t} + \alpha \frac{\partial{\beta}}{\partial t}+ \sigma \frac{\partial{s}}{\partial t}
+\varepsilon (\rho,s)+\Psi+ \frac{1}{2}(\vec \nabla \nu + \alpha \vec \nabla \beta + \sigma \vec \nabla s )^2]
\nonumber \\
{\cal L}_{\vec v} &\equiv & \frac{1}{2}\rho (\vec v-\hat{\vec v})^2
\nonumber \\
{\cal L}_{boundary} &\equiv & \frac{\partial{(\nu \rho)}}{\partial t} + \vec \nabla \cdot (\nu \rho \vec v )
\label{Lagactionsimp5b}
\enr
The only term containing $\vec v$ is ${\cal L}_{\vec v}$, it can easily be seen that
this term will lead, after we nullify the variational derivative, to \ern{vformb} but will otherwise
have no contribution to other variational derivatives. Notice that the term ${\cal L}_{boundary}$
contains only complete partial derivatives and thus can not contribute to the equations although
it can change the boundary conditions. Hence we see that \ers{lagmulb}, \ern{nueqb}, \ern{sigmaeq} and \ern{alphacon}
can be derived using the Lagrangian density $\hat {\cal L}$ in which $\hat{\vec v}$ replaces
$\vec v$ in the relevant equations. Furthermore, after integrating the six \eqs
(\ref{lagmulb},\ref{nueqb},\ref{sigmaeq},\ref{alphacon}) we can insert the functions $\alpha,\beta,\nu,\sigma,s$
into \ern{vformb} to obtain the physical velocity $\vec v$.
Hence, the general non-barotropic fluid dynamics problem is changed such that instead of
solving the five equations
(\ref{massconb},\ref{Eulerb},\ref{seq}) we need to solve an alternative set of six equations which can be derived from the Lagrangian density $\hat {\cal L}$.
We notice that the potential $\Psi$ affects the flow dynamics through a single equation (\ref{nueqb}) and
thermodynamics affect the flow dynamics through two equations (\ref{nueqb}) which depends on the specific enthalpy $w$ and (\ref{sigmaeq}) which depends on the temperature $T$.

\section{Stationary fluid dynamics}
\label{Statflow}

Stationary flows are a unique phenomena of Eulerian fluid dynamics which has
no counter part in Lagrangian fluid dynamics. The stationary flow is defined
by the fact that the physical fields $\vec v,\rho,s$ do not depend on the
temporal coordinate. This however does not imply that the stationary potentials
$\alpha,\beta,\nu,\sigma$ are all functions of spatial coordinates alone.
Moreover, it can be shown that choosing the potentials in such a way will lead
to erroneous results in the sense that the stationary equations of motion can
not be derived from the Lagrangian density $\hat {\cal L}$ given in \ern{Lagactionsimp5b}.
However, this problem can be amended easily
as follows. Let us choose $\alpha,\nu,\sigma$ to depend on the spatial coordinates alone.
Let us choose $\beta$ such that:
\beq
\beta = \bar \beta - t
\label{betastas}
\enq
in which $\bar \beta$ is a function of the spatial coordinates. The Lagrangian
density $\hat {\cal L}$ given in \ern{Lagactionsimp5b} will take the form:
\beq
\hat {\cal L} = \rho \left(\alpha -\varepsilon (\rho,s)-\Psi -
\frac{1}{2}(\vec \nabla \nu + \alpha \vec \nabla \beta+ \sigma \vec \nabla s)^2\right)
\label{stathatLb}
\enq
Varying the Lagrangian $\hat {L} = \int \hat {\cal L} d^3x$ with respect to $\alpha,\beta,\nu,\rho,\sigma,s$ leads to the following equations:
\ber
& &  \vec \nabla \cdot (\rho \hat {\vec v} ) = 0
\nonumber \\
& & \rho \hat {\vec v} \cdot \vec \nabla \alpha = 0
\nonumber \\
& & \rho (\hat {\vec v} \cdot \vec \nabla \bar \beta - 1) = 0
\nonumber \\
& & \alpha= \frac{1}{2} \hat {\vec v}^2 + w + \Psi
\nonumber \\
& & \rho \hat {\vec v} \cdot \vec \nabla s = 0
\nonumber \\
& & \rho \hat {\vec v} \cdot \vec \nabla \sigma = \rho T
\label{statlagmulb}
\enr
Which simplifies for any spatial point in which the fluid density is not null to:
\ber
& &  \vec \nabla \cdot (\rho \hat {\vec v} ) = 0
\nonumber \\
& &  \hat {\vec v} \cdot \vec \nabla \alpha = 0
\nonumber \\
& &  \hat {\vec v} \cdot \vec \nabla \bar \beta = 1
\nonumber \\
& & \alpha= \frac{1}{2} \hat {\vec v}^2 + w + \Psi
\nonumber \\
& &  \hat {\vec v} \cdot \vec \nabla s = 0
\nonumber \\
& & \hat {\vec v} \cdot \vec \nabla \sigma =  T
\label{statlagmulb2}
\enr
$\alpha$ is thus the Bernoulli constant (this was also noticed in \cite{Yahalom}).
Calculations similar to the ones done in previous subsections will show that those equations
lead to the stationary Euler equations:
\beq
(\vec v \cdot \vec \nabla) \vec v = -\frac{\vec \nabla p}{\rho }  -  \vec \nabla \Psi
\label{Eulerstatb}
\enq

\section {Constants of Motion}
\label{Topflu}

A well known concept in fluid dynamics is circulation around a trajectory that is comoving with the
flow:
\beq
C \equiv \oint \vec v \cdot d \vec l
\label{circ}
\enq
$d \vec l$ is infinitesimal length oriented along the trajectory. It follows from \ern{dvform2bb}:
that:
\beq
\frac{d C}{d t} = \oint T \vec \nabla s \cdot d \vec l =  \oint T d s
\label{dcircdt}
\enq
hence circulation is not conserved unless the flow is barotropic (which means a uniform specific entropy throughout the flow), or the trajectory happens to be on a constant specific entropy or constant temperature surface. However, using the variational equations that we obtained
in the previous sections we can generalized circulations conservation to an arbitrary comoving
trajectory as follows. First we define a topological velocity field:
\beq
\vec v_t \equiv \vec v - \sigma \vec \nabla s.
\label{vt}
\enq
$\sigma$ and $s$ are defined in previous sections. Next we define a circulation of the topological flow field in an analogue way to the definition of standard circulation:
\beq
C_t \equiv \oint \vec v_t \cdot d \vec l = C - \oint \sigma \vec \nabla s \cdot d \vec l
= C - \oint \sigma d s.
\label{circt}
\enq
It is easy to show that this quantity is conserved for any trajectory:
\beq
\frac{d C_t}{d t} = \frac{d C}{d t} - \oint \frac{d  \sigma }{d t}d s =
 \oint T d s - \oint T d s = 0.
\label{dcirctdt}
\enq
in which we have used \ern{sigmaeq} and \ern{dcircdt}. Taking into account the
generalized Clebsch representation \ern{vformb} it follows that:
\beq
\vec v_t = \vec v -  \sigma \vec \nabla s = \alpha \vec \nabla \beta + \vec \nabla \nu.
\label{vCt}
\enq
Hence the topological flow field has a standard Clebsch form. Moreover, the associated
vorticity of this flow is:
\beq
\vec \omega_t = \vec \nabla \times \vec v_t =  \vec \nabla \alpha \times \vec \nabla \beta
= \vec \omega -  \vec \nabla \sigma \times \vec \nabla s,
\label{vorticb2}
\enq
We can use Stokes theorem to write the topological circulation $C_t$ as topological vorticity flux
$\Phi_t$ through a comoving surface:
\beq
C_t = \oint \vec v_t \cdot d \vec l = \int \vec \nabla \times \vec v_t \cdot d \vec A
= \int \vec \omega_t \cdot d \vec A =
 \int \vec \nabla \alpha \times \vec \nabla \beta \cdot d \vec A = \int d \alpha d \beta =
\Phi_t.
\label{circt2}
\enq
Hence a tube of topological vorticity lines will not change its flux due to the flow dynamics. This is of course true even in the cross section of the tube is infinitesimal, which means that each vortex line is comoving and thus the topology of the topological vortex lines is conserved as well, and they cannot be unknotted by the flow if they are initially knotted. This is of course the reason
why they are called topological in the first place. It is easy to show using \ern{lagmulb} and \ern{alphacon} that:
\beq
\frac{\partial{\vec \omega_t}}{\partial t} = \vec \nabla \times (\vec v \times \vec \omega_t).
\label{omegteq1}
\enq
which is a sufficient condition for the topological vortex line to co-move with the velocity field $\vec v$. Barotropic fluid dynamics is known to have the helicity topological constant of motion;
\beq
{\cal H} \equiv \int \vec \omega \cdot \vec v d^3 x,
\label{hel}
\enq
which is known to measure the degree of knottiness of lines of the vorticity field $\vec \omega$ \cite{Moff2}. However, this quantity is not conserved in non-barotropic flows. Nevertheless, the previous discussion shows that this quantity can be easily generalized using the concept of topological flow fields and their topological vorticity:
\beq
{\cal H}_t \equiv \int \vec \omega_t \cdot \vec v_t d^3 x.
\label{heltop}
\enq
A straightforward calculation will show that this is indeed conserved.

Let us write the topological constants given in \ern{heltop}  in terms
of the fluid dynamical potentials $\alpha, \beta, \nu$ introduced
in previous sections, the scalar product $\vec \omega_t \cdot \vec v_t$ is thus:
\beq
\vec \omega_t \cdot \vec v_t =  (\vec \nabla \alpha \times \vec \nabla \beta) \cdot \vec \nabla \nu.
\label{vecpotflu4}
\enq
We introduce local vector basis: $(\vec \nabla \alpha, \vec \nabla \beta, \vec \nabla \mu)$
in which $\mu$ is a comoving function with a gradient which is not parallel
to $\vec \nabla \alpha, \vec \nabla \beta$. Using those functions we can write $\vec \nabla \nu$ as:
\beq
\vec \nabla \nu = \frac{\partial \nu}{\partial \alpha} \vec \nabla \alpha +  \frac{\partial \nu}{\partial \beta} \vec \nabla \beta+ \frac{\partial \nu}{\partial \mu} \vec \nabla \mu .
\label{nudiv}
\enq
Hence we can write:
\beq
\vec \omega_t \cdot \vec v_t = \frac{\partial \nu}{\partial \mu}
(\vec \nabla \alpha \times \vec \nabla \beta) \cdot  \vec \nabla \mu =
\frac{\partial \nu}{\partial \mu} \frac{\partial(\alpha,\beta,\mu)}{\partial(x,y,z)} .
\label{vecpotflu5}
\enq
Now we can insert \er{vecpotflu5} into \er{heltop} to obtain the expression:
\beq
{\cal H}_t = \int \frac{\partial \nu}{\partial \mu} d \mu  d\alpha d\beta.
\label{hel2}
\enq
The reader should notice that in some scenarios it may be that the flow domain should be divided into patches
in which different definitions of $\mu,\alpha,\beta$ apply to different domains, we do not see this
as a limitation for our formalism since the topology of the flow is conserved by the flow equations.
In those cases ${\cal H}_t$ should be calculated as sum of the contributions from each patch.
We can think about the fluid domain as composed of thin closed tubes of
topological vortex lines each labelled by $(\alpha, \beta)$. Performing the integration along such a thin tube in the metage direction results in:
\beq
\oint_{\alpha, \beta} \frac{\partial \nu}{\partial \mu} d \mu = [\nu]_{\alpha, \beta},
\label{nucutflu}
\enq
in which $[\nu]_{\alpha, \beta}$ is the discontinuity of the function $\nu$ along
its cut. Thus a thin tube of vortex lines in which $\nu$ is single valued does not
contribute to the topological helicity integral. Inserting \ern{nucutflu} into \ern{hel2}
will result in:
\beq
{\cal H}_t = \int [\nu]_{\alpha, \beta}  d\alpha d\beta= \int [\nu] d\Phi_t,
\label{hel3}
\enq
in which we have used \ern{circt}. Hence:
\beq
[\nu] = \frac{d{\cal H}_t}{d\Phi_t},
\enq
the discontinuity of $\nu$ is thus the density of topological helicity per unit of topological vortex flux in a tube.
We deduce that the Clebsch representation does not entail zero topological helicity, rather it is
perfectly consistent with non zero topological helicity as was demonstrated above. Further more according to \er{nueqb}
\beq
\frac{d [\nu]}{d t} = 0.
\label{nucuteqflu}
\enq
We conclude that not only is the topological helicity conserved as an integral
quantity of the entire flow domain but also the (local) density of topological helicity per
unit of topological vortex flux is a conserved quantity as well.

\section{A simpler variational principle of non-stationary fluid dynamics}

Lynden-Bell \& Katz \cite{LynanKatz} have shown that an Eulerian variational principle for
non-stationary barotropic fluid dynamics can be given in terms of two functions the density $\rho$ and the load $\lambda$. However, their velocity was given an
implicit definition in terms of a partial differential equation and its variations was constrained
to satisfy this equation. Much the same criticism holds for their general variational for non-barotropic flows \cite{KatzLyndeb}. However, Yahalom \& Lynden-Bell \cite{AYLB}
have shown that a true variational principle (that is unconstrained and without implicit definitions) for barotropic flows can be given in terms of three functions $\rho, \nu$ and
an additional comoving function $\lambda$. Below we shall show that for the non-barotropic
case four functions will suffice those are $\rho, \nu, s$ and an additional comoving function.

Consider the last two equations in (\ref{lagmulb}) and write them explicitly
in term of the generalized Clebsch form given in \ern{vformb}, we obtain:
\ber
& & \frac{d \beta}{dt} = \frac{\partial \beta}{\partial t} + \vec v \cdot \vec \nabla \beta
= \frac{\partial \beta}{\partial t} + (\alpha \vec \nabla \beta + \sigma \vec \nabla s + \vec \nabla \nu) \cdot \vec \nabla \beta
 = 0
\nonumber \\
& & \frac{d s}{dt} = \frac{\partial s}{\partial t} + \vec v \cdot \vec \nabla s
= \frac{\partial s}{\partial t} + (\alpha \vec \nabla \beta + \sigma \vec \nabla s + \vec \nabla \nu) \cdot \vec \nabla s  = 0.
\label{lagmulb3}
\enr
Those are two algebraic equations for the variables $\alpha,\sigma$ which can be readily
solved. Introducing the notation:
\ber
& & a_{\beta \beta} = (\vec \nabla \beta)^2, \qquad a_{s s} = (\vec \nabla s)^2, \qquad
a_{s \beta} = a_{\beta s} = \vec \nabla \beta \cdot \vec \nabla s
\nonumber \\
& & k_\beta= - \frac{\partial \beta}{\partial t} - \vec \nabla \nu \cdot \vec \nabla \beta, \qquad
k_s= - \frac{\partial s}{\partial t} - \vec \nabla \nu \cdot \vec \nabla s.
\label{akdef}
\enr
We obtain both $\alpha,\sigma$ as functionals of the variables $s,\beta,\nu$:
\beq
\alpha[s,\beta,\nu] =
 \frac{a_{s s} k_\beta - a_{s \beta} k_s}{a_{s s} a_{\beta \beta} - a_{s \beta}^2}, \qquad
 \sigma[s,\beta,\nu] =
 \frac{a_{\beta \beta} k_s - a_{s \beta} k_\beta}{a_{s s} a_{\beta \beta} - a_{s \beta}^2}
 \label{alsig}
\enq
 Similarly the velocity field is now a functional of the three variables $s,\beta,\nu$:
 \beq
\vec v [s,\beta,\nu] = \alpha [s,\beta,\nu] \vec \nabla \beta + \sigma [s,\beta,\nu] \vec \nabla s + \vec \nabla \nu
\label{vformbthree}
\enq
Finally the Lagrangian density is a functional of the four variables $s,\beta,\nu,\rho$:
\ber
\hat {\cal L} [s,\beta,\nu,\rho] &=&  -\rho \left[\frac{\partial{\nu}}{\partial t} + \alpha [s,\beta,\nu] \frac{\partial{\beta}}{\partial t}+ \sigma [s,\beta,\nu] \frac{\partial{s}}{\partial t} +\varepsilon (\rho,s)+\Psi \right.
\nonumber \\
&+& \left. \frac{1}{2}(\vec \nabla \nu + \alpha [s,\beta,\nu] \vec \nabla \beta + \sigma [s,\beta,\nu] \vec \nabla s )^2\right]
\label{Lagactionsimp9}
\enr
The variation of which will lead to the following four equations which replace
the original set of equation (\ref{massconb},\ref{Eulerb},\ref{seq}):
\beq
\frac{\partial{\rho}}{\partial t} + \vec \nabla \cdot (\rho \vec v ) = 0, \quad
\frac{d \nu}{d t} = \frac{1}{2} \vec v^2 - w - \Psi, \quad
\frac{d \sigma [s,\beta,\nu]}{dt} = T, \quad
\frac{d \alpha [s,\beta,\nu]}{dt} = 0.
\label{fourreeq}
\enq
Written explicitly the form of those equations may look rather complicated.

\section {Conclusion}

In this paper we have discussed the variational analysis of non-barotropic flows starting from
a classical Lagrangian variational analysis, and moving later to an Eulerian variational analysis
which demanded auxiliary functions. This lagrangian which depends on nine functions: the original set of five functions $\rho,\vec v,s$ and auxiliary variables $\alpha, \beta, \nu, \sigma$ was reduces to a six function Lagrangian depending on  $\alpha, \beta, \nu, \sigma,\rho, s$. This was
further reduced to a four function Lagrangian of the remaining variables $ \beta, \nu, \rho, s$,
leading to a set of four cumbersome equations. We have also dedicated a paragraph to the stationary
version of our Lagrangian formalism. Our result can be compared with respect to what was achieved in the last realistic barotropic case, there a three function formalism is available for both stationary and non-stationary flows. In both barotropic and non-barotropic cases variational formalism has reduced the number of variables needed with respect to the physical description.
However, this economy of variables comes with a topological cost. In the barotropic case the vorticity lines must lie on surfaces and cannot be volume filling. This is also true for non-barotropic flows in which one demands the same for the topological vorticity.

The problem of stability analysis and the description of numerical schem\-es using the described variational principles exceed the scope of this paper.
We suspect that for achieving this we will need to add additional
constants of motion constraints to the action as was done by \cite{Arnold1,Arnold2}
see also \cite{YahalomKatz,YahalomMonth}. Additional points for future study include the Noether currents
of the present variational formalism and their implications. Also \cite{Threefunmhd} suggests that as in the magnetohydrodynamic case there may be a way to reduce the number of variables
to three for non-barotropic {\bf stationary} fluid dynamics. Hopefully this will be discussed in a future paper.

\appendix
\setcounter{section}{0}
 \section{Variational Formalism of Point Particles}
\label{pointpart}

 Consider a classical particle with the coordinates $\vec x (t)$ and mass $m$ interacting with a scalar potential $\Psi (\vec x,t)$.
 We will not be interested in the effects of the particle on the field and thus consider the field as "external". The action of the said particle is:
 \ber
 {\cal A} &=& \int_{t1}^{t2} L dt,
 \nonumber \\
 L &\equiv& \frac{1}{2} m v^2 - m \Psi,  \qquad
  \vec v \equiv \frac{d \vec x}{dt} \equiv \dot{\vec x}, \quad v =|\vec v|.
 \label{classparticleact}
 \enr
The variation of the Lagrangian is given by:
\beq
 \delta L =  m \dot{\vec x  } \cdot \delta \dot{\vec x} - m \vec \nabla \Psi \cdot \delta \vec x =
 \frac{d (m \vec v \cdot \delta \vec x)}{dt} - m \dot {\vec v} \cdot \delta \vec x
 - m \vec \nabla \Psi \cdot \delta \vec x
 \label{delLf}
 \enq
  \beq
  \delta L =
   \frac{d \left( m \vec v \cdot \delta \vec x\right)}{dt} +(-m \dot{\vec v} +\vec F) \cdot \delta \vec x. \qquad \vec F \equiv - m \vec \nabla \Psi.
 \label{delL1}
 \enq
 Thus the variation of the action is:
  \beq
  \delta A =\int_{t1}^{t2} \delta L dt =
  \left. m \vec v  \cdot \delta \vec x \right|_{t1}^{t2}
   - \int_{t1}^{t2}(m \dot{\vec v} - \vec F) \cdot \delta \vec x dt.
 \label{delA1}
 \enq
 Since the classical trajectory is such that the variation of the action on it vanishes for
 a small modification of the trajectory $\delta \vec x$ that vanishes at $t1$ and $t2$ but is otherwise arbitrary it follows that:
 \beq
  m \dot{\vec v} = \vec F.
 \label{equamotion}
 \enq
  For a system of $N$ particles each with an index $1 \leq j \leq N$, a corresponding mass $m_j$, charge  $e_j$, position vector $\vec x_j$ and velocity $\vec v_j \equiv \frac{d \vec x_j}{dt}$ the action and Lagrangian for each point particle are as follows:
 \ber
 {\cal A}_j &=& \int_{t1}^{t2} L_j dt,
 \nonumber \\
 L_{j} &\equiv& \frac{1}{2} m_j v_j^2 - m_j \Psi_j.
 \label{classparticlej}
 \enr
 The action and Lagrangian of the system of particles is:
\beq
 {\cal A}_s = \int_{t1}^{t2} L_s dt, \qquad L_s = \sum_{j=1}^{N} L_j.
  \label{classparticlesys}
 \enq
 The variational analysis follows the same lines as for a single particle and we obtain a set of equations of the form:
 \beq
 m_j \dot{\vec v}_j = \vec F_j , \qquad j \in [1-N].
 \label{equamotionj}
 \enq

\PublishersNote{}


\begin {thebibliography}{9}
\bibitem{AYLB}
Asher Yahalom and Donald Lynden-Bell "Variational Principles for Topological Barotropic Fluid Dynamics" ["Simplified Variational Principles for Barotropic Fluid Dynamics" Los-Alamos Archives - physics/ 0603162] Geophysical \& Astrophysical Fluid Dynamics. 11/2014; 108(6). DOI: 10.1080/03091929.2014.952725.
\bibitem{Moff2}
H.K. Moffatt, \index{Moffatt H.K.} "The degree of knottedness of tangled vortex lines," {\it J. Fluid Mech.}, vol. 35, 117, 1969.
\bibitem {Eckart}
C. Eckart\index{Eckart C.}, "Variation\index{variational
principle} Principles of Hydrodynamics \index{hydrodynamics},"
{\it Phys. Fluids}, vol. 3, 421, 1960.
\bibitem {Bertherton}
F.P. Bretherton "A note on Hamilton's principle for perfect fluids," Journal of Fluid Mechanics / Volume 44 / Issue 01 / October 1970, pp 19 31 DOI: 10.1017/S0022112070001660, Published online: 29 March 2006.
\bibitem {Clebsch1}
Clebsch, A. 1857 Uber eine allgemeine transformation der hydro-dynamischen Gleichungen. J. Reine Angew. Math. 54, 293-312.
\bibitem {Clebsch2}
Clebsch, A. 1859 Uber die Integration der hydrodynamischen Gleichungen. J. Reine Angew. Math. 56, 1-10.
\bibitem {Davidov}
B. Davydov\index{Davydov B.}, "Variational principle and canonical
equations for an ideal fluid," {\it Doklady Akad. Nauk},  vol. 69,
165-168, 1949. (in Russian)
\bibitem{Seliger}
R. L. Seliger  \& G. B. Whitham, {\it Proc. Roy. Soc. London}, A{\bf 305}, 1 (1968)
\bibitem{nonBarotropic}
A. Yahalom "Simplified Variational Principles for non-Barotropic
Magnetohydrodynamics". (arXiv: 1510.00637 [Plasma Physics]). J. Plasma Phys. vol. 82, 905820204 2016.  doi:10.1017/S0022377816000222.
\bibitem{nonBarotropic2}
 A. Yahalom "Non-Barotropic Magnetohydrodynamics as a Five Function Field Theory". International Journal of Geometric Methods in Modern Physics, No. 10 (November 2016). Vol. 13 1650130 (13 pages) \copyright World Scientific Publishing Company, DOI: 10.1142/S0219887816501309.
 \bibitem{Lamb H.}
H. Lamb {\it Hydrodynamics} Dover Publications (1945).
\bibitem{Yahalom}
A. Yahalom, "Method and System for Numerical Simulation of Fluid Flow", US patent 6,516,292 (2003).
\bibitem{LynanKatz}
D. Lynden-Bell and J. Katz "Isocirculational Flows and their Lagrangian and Energy principles",
Proceedings of the Royal Society of London. Series A, Mathematical and Physical Sciences, Vol. 378,
No. 1773, 179-205 (Oct. 8, 1981).
\bibitem{KatzLyndeb}
J. Katz \& D. Lynden-Bell 1982,{\it Proc. R. Soc. Lond.} {\bf A 381} 263-274.
\bibitem{Arnold1}
V. I. Arnold "A variational principle for three-dimensional steady flows of an ideal fluid",
Appl. Math. Mech. {\bf 29}, 5, 154-163.
\bibitem{Arnold2}
V. I. Arnold "On the conditions of nonlinear stability of planar curvilinear flows of an ideal fluid", Dokl. Acad. Nauk SSSR {\bf 162} no. 5.
\bibitem{YahalomKatz}
Yahalom A., Katz J. \& Inagaki K. 1994, {\it Mon. Not. R. Astron. Soc.} {\bf 268} 506-516.
fluid", Dokl. Acad. Nauk SSSR {\bf 162} no. 5.
\bibitem {YahalomMonth}
    \textsc{Yahalom A.} 2011 Stability in the Weak Variational Principle of Barotropic Flows and Implications for Self-Gravitating Discs
    \emph{Monthly Notices of the Royal Astronomical Society} \textbf{418},
     401--426.
\bibitem{Threefunmhd}
Yahalom, Asher. 2021. "A Three-Function Variational Principle for Stationary Nonbarotropic Magnetohydrodynamics" Symmetry 13, no. 9: 1632. https://doi.org/10.3390/sym13091632

\end {thebibliography}
\end{document}